\documentclass[final,11pt,2p,times,sort&compress]{elsarticle} 
  
  \usepackage{color}
  \usepackage{amsmath}
  \usepackage{calligra}
  \usepackage{caption}
  \usepackage{cprotect}
  \usepackage{enumerate}
  \usepackage[T1]{fontenc}
  \usepackage{ulem}
  \usepackage{lineno}
  \usepackage{multicol}
  \usepackage{tabularx}
  \usepackage[]{titlesec}
  \newcolumntype{L}[1]{>{\raggedright\arraybackslash}p{#1} }
  \newcolumntype{C}[1]{>{\centering  \arraybackslash}p{#1} }
  \newcolumntype{R}[1]{>{\raggedleft \arraybackslash}p{#1} }
\journal{Computational Materials Science}
\begin{document}
\begin{frontmatter}
  \title{The AFLOW Standard for High-Throughput Materials Science Calculations}
  \author{Camilo E. Calderon$^1$, Jose J. Plata$^1$, Cormac Toher$^1$, Corey Oses$^1$, Ohad Levy$^{1,\dagger}$, \\
          Marco Fornari$^2$, Amir Natan$^3$, Michael J. Mehl$^4$, Gus Hart$^5$, Marco Buongiorno Nardelli$^6$, Stefano Curtarolo$^{7,\star}$}
  \address{$^{1}$Department of Mechanical Engineering and Materials Science, 
                 Duke University, Durham, North Carolina 27708, USA}
  \address{$^{2}$Department of Physics, Central Michigan University, Mount Pleasant, MI 48858, USA }
  \address{$^{3}$ Department Of Physics And Electrical Engineering \& Electronics, Faculty of Engineering, Tel Aviv University, Tel Aviv 69978, Israel }
  \address{$^{4}$ Center for Computational Materials Science, Naval Research Laboratory, Washington, DC 20375-5345, USA}
  \address{$^{5}$ Department of Physics and Astronomy, Brigham Young University, Provo, Utah 84602, USA}
  \address{$^{6}$Department of Physics and Department of Chemistry, University of North Texas, Denton TX }
  \address{$^{7}$Materials Science, Electrical Engineering, Physics and Chemistry, Duke University, Durham NC, 27708}
  \address{$^{\dagger}$On leave from the Physics Department, NRCN, Israel}
  \address{$^{\star}${\bf corresponding:} stefano@duke.edu}
  \begin{abstract}
    The Automatic-Flow ({\small AFLOW}) standard for the high-throughput construction of materials science 
    electronic structure databases is described. Electronic structure calculations of solid state materials depend
    on a large number of parameters which must be understood by
    researchers, and must be reported by originators to ensure reproducibility and enable collaborative
  database expansion. We therefore describe standard parameter values for {\textit k}-point grid density, 
    basis set plane wave kinetic energy cut-off, exchange-correlation 
    functionals, pseudopotentials, DFT+U parameters, and convergence criteria used in {\small AFLOW} calculations.
  \end{abstract}
\end{frontmatter}

\section{Introduction} \label{sec:Intro}

The emergence of computational materials science over the last two decades has been inextricably linked to the
development of complex quantum-mechanical codes that enable accurate evaluation of the electronic and 
thermodynamic properties of a wide range of materials. The continued advancement of this field entails the 
construction of large open databases of materials properties that can be easily reproduced and extended.
One obstacle to the reproducibility of the data is the unavoidable complexity of the codes used to obtain 
it. Published data usually includes basic information about the underlying calculations that  allows rough 
reproduction. However, exact duplication depends on many details, that are seldom reported, and is therefore 
difficult to achieve.
  
These difficulties might limit the utility of the databases currently being created by high-throughput frameworks, 
such as {\small AFLOW}\cite{aflowPAPER, aflowlibPAPER, aflowAPI} and the Materials Project
\cite{APL_Mater_Jain2013,CMS_Ong2012b}. For maximal 
impact, the data stored in these repositories must be generated and represented in a consistent and robust manner, 
and shared through standardized calculation and communication protocols. Following these guidelines would promote 
optimal use of the results generated by the entire community.

The {\small AFLOW} (Automatic FLOW) code is a framework for high-throughput computational materials discovery
\cite{aflowPAPER, aflowlibPAPER, aflowAPI, aflowlib}, using separate DFT packages to calculate electronic
structure and optimize the atomic geometry. The {\small AFLOW} framework works with the 
{\small VASP}\cite{VASP4_1, VASP4_2, VASP4_3, VASP4_4} DFT package,
and integration with the {\small Quantum ESPRESSO} software
\cite{quantum_espresso_2009} is currently in progress. The {\small AFLOW} framework includes 
preprocessing functions for generating input files for the DFT package; obtaining the initial geometric structures 
by extracting the relevant data from crystallographic information files or by generating them using inbuilt prototype
databases, and then transforming them into standard forms which are easiest to calculate. It then runs and monitors 
the DFT calculations automatically, detecting and responding to calculation failures, whether they are due to insufficient
hardware resources or to runtime errors of the DFT calculation itself. Finally, {\small AFLOW} contains postprocessing 
routines to extract specific properties from the results of one or more of the DFT calculations, such as the band 
structure or thermal properties \cite{Toher_PRB_AGL_2014}.

The {\small AFLOWLIB} repository \cite{aflowlibPAPER, aflowAPI,
  aflowlib} was built according to these principles of consistency and reproducibility,
and the data it contains can be easily accessed through a representational state transfer application programming 
interface (REST-API) \cite{aflowAPI}. In this paper we present a detailed description of the {\small AFLOW} standard 
for high-throughput (HT) materials science calculations by which the data in this repository was created. 

\section{{\small AFLOW} Calculation Types} \label{sec:AFLOWtypes}

The {\small AFLOWLIB} consortium\cite{aflowlibPAPER} repository is divided into databases containing calculated 
properties of over 625,000 materials: 
the Binary Alloy Project, the Electronic Structure database, the Heusler database, and the Elements database. 
These are freely accessible online via the {\small AFLOWLIB} website\cite{aflowlib}, as well as through the 
API\cite{aflowAPI}. The Electronic Structure database consists of entries found in the Inorganic Crystal 
Structures Database, ICSD\cite{ICSD, ICSD3}, and will thus be referred to as ``ICSD'' throughout this publication.
The Heusler database consists of ternary compounds, primarily based on the Heusler structure but with other
structure types now being added.

The high-throughput construction of these materials databases relies on a pre-defined set of standard {\textit{calculation 
types}}. These are designed to accommodate the interest in various properties of a given material (e.g. the ground 
state ionic configuration, thermodynamic quantities, electronic and
magnetic properties), the program flow of the HT framework that
envelopes the DFT portions of the calculations, as well as the practical 
need for computational robustness. The {\small AFLOW} standard thus deals with the parameters involved in the following 
calculation types:

\begin{enumerate}[i.]
  \item {{\verb!RELAX!}.} Geometry optimizations using algorithms implemented within the DFT package. This calculation
  type is concerned with obtaining the ionic configuration and cell
  shape and volume that correspond to a minimum in the 
  total energy. It consists of two sequential relaxation steps. The starting point for the first step, {\verb!RELAX1!}, 
  can be an entry taken from an external source, such as a library of alloy 
  prototypes\cite{Massalski, curtarolo:calphad_2005_monster}, the ICSD database, or the Pauling 
  File\cite{PaulingFile}. These initial entries are preprocessed by 
  {\small AFLOW}, and cast into a unit cell that is most convenient
  for calculation, usually the standard primitive cell, in the format appropriate for the DFT package in use. The second step, {\verb!RELAX2!}, 
  uses the final ionic positions from the first step as its starting point, and serves as a type of annealing step. 
  This is used for jumping out of possible local minima resulting from wavefunction artifacts.
  \item {{\verb!STATIC!}.} A single-point energy calculation. The starting point is the set of final ionic positions, 
  as produced by the {\verb!RELAX2!} step. The outcome of this calculation is used in the determination of most 
  of the thermodynamic and electronic properties included in the various {\small AFLOW} databases.
 It therefore applies a more demanding set of parameters than those used on the {\verb!RELAX!} 
  set of runs.
  \item {{\verb!BANDS!}.} Electronic band structure generation. The converged {\verb!STATIC!} charge 
  density and ionic positions are used as the starting points, and the wavefunctions are reoptimized along standardized 
  high symmetry lines connecting special {\textit{k}}-points in the irreducible Brillouin zone (IBZ)\cite{curtarolo:art58}.
\end{enumerate}

These calculation types are performed in the order shown above (i.e. {\verb!RELAX1!} $\rightarrow$ {\verb!RELAX2!} 
$\rightarrow$ {\verb!STATIC!} $\rightarrow$ {\verb!BANDS!}) on all materials found in the Elements, 
ICSD, and Heusler databases. Those found in the Binary Alloy database contain data produced only by the two 
{\verb!RELAX!} calculations. 
Sets of these calculation types can be combined to describe more complex
phenomena than can be obtained from a single calculation. For
example, sets of {\verb!RELAX!} and  {\verb!STATIC!} calculations for different cell
volumes and/or atomic configurations are used to calculate
thermal and mechanical properties by the {\small AGL}
\cite{Toher_PRB_AGL_2014} and {\small APL} \cite{aflowPAPER}  methods
inplemented within the {\small AFLOW} framework. 
In the following, we describe the parameter sets used to address the
particular challenges of the calculations included in each {\small AFLOW} repository. 

\section{The {\small AFLOW} standard parameter set} \label{sec:AFLOWstandard}

The standard parameters described in this work are classified according to the wide variety of tasks that a typical solid 
state DFT calculation involves: Brillouin zone sampling, Fourier transform meshes, basis sets, potentials, 
self-interaction error (SIE) corrections, electron spin, algorithms guiding SCF convergence and ionic relaxation, and
output options.

Due to the intrinsic complexity of the DFT codes it is impractical to
specify the full set of DFT calculation parameters within an HT framework. Therefore, the {\small AFLOW} standard 
adopts many, but not all, of the internal defaults set by the DFT software package. This is most notable in the description of the
Fourier transform meshes, which rely on a discretization scheme that depends on the applied basis and crystal 
geometry for its specification. Those internal default settings are cast aside when 
error corrections of failed DFT runs, an integral part of {\small AFLOW}'s functionality, take place. The settings 
described in this work are nevertheless prescribed as fully as is practicable, in the interest of providing as 
much information as possible to anyone interested in reproducing or building on our results.

\subsection{ {\textit k}-point sampling} \label{subsec:kpointgrid}

Two approaches are used when sampling the IBZ: the first consists of uniformly distributing a large number 
of {\textit k}-points in the IBZ, while the second relies on the construction of paths connecting high symmetry (special) 
{\textit k}-points in the IBZ. Within {\small AFLOW}, the second sampling method corresponds to the {\verb!BANDS!} 
calculation type, whereas the other calculation types (non-{\verb!BANDS!}) are performed using the first sampling 
method.

Sampling in non-{\verb!BANDS!} calculations is obtained by defining and setting $N_{KPPRA}$, the number of 
{\textit k}-points per atom. This quantity determines the total number of {\textit k}-points in the IBZ, 
taking into account the {\textit k}-points density along each reciprocal lattice vector as well as the number of atoms 
in the simulation cell, via the relation:

\begin{equation} \label{eq:kppra}
  { N_{KPPRA} \leq min \left[ \prod\limits_{i=1}^3 N_i \right] \times N_{\text{a} } }
\end{equation}

\noindent $N_{\text{a}}$ is the number of atoms in the cell, and the $N_{i}$ factors correspond to  the number 
of sampling points along each reciprocal lattice vector,
$\vec{b_{i}}$, respectively. These factors define the grid resolution, 
${\it \delta} k{_i} {\| \vec{b{_i}} \|}/{N{_i} }$, which is made as uniform as possible 
under the constraint of Eqn. \ref{eq:kppra}. The {\textit k}-point meshes are then 
generated within the Monkhorst-Pack scheme\cite{Monkhorst1976}, unless the material belongs to the
{\textit{hP}}, or {\textit{hR}} Bravais lattices, in which case the hexagonal symmetry is preserved by centering the mesh 
at the $\Gamma$-point.

Default $N_{KPPRA}$ values depend on the calculation type and the
database. The $N_{KPPRA}$ values used for the entries in the Elements
database are material specific and set manually due to convergence of
the total energy calculation. The defaults applied to the 
{\verb!RELAX!} and {\verb!STATIC!} calculations are summarized in
Table \ref{tab:kgridnonbands}.
These defaults ensure proper convergence of the calculations. They
may be too stringent for some cases but enable reliable
application within the HT framework, thus presenting a practicable
balance between accuracy and calculation cost.

{\small
   \begin{table}[h]
     \centering
     \begin{tabular}[b]{L{2.5cm} R{2.5cm} R{2.5cm}}
       \hline
       \hline
       Database      & {\verb!STATIC!} & {\verb!RELAX!} \\
       \hline
       Binary Alloy  &  N.A.           &  6000          \\
       Heusler       & 10000           &  6000          \\
       ICSD          & 10000           &  8000          \\
       \hline
       \hline
     \end{tabular}
     \cprotect\caption{Default $N_{KPPRA}$ values used in non-{\verb!BANDS!} calculations.}
     \label{tab:kgridnonbands}
   \end{table}
}

For {\verb!BANDS!} calculations {\small AFLOW}  generates Brillouin zone integration 
paths in the manner described in a previous
publication\cite{curtarolo:art58}. The \textit{k}-point sampling density is the {\textit{line 
density}} of {\textit k}-points along each of the straight-line
segments of the path in the IBZ. The default setting
of {\small AFLOW} is 128 {\textit k}-points along each segment connecting high-symmetry {\textit k}-points in 
the IBZ for single element structures, and 20 {\textit
  k}-points for compounds. 

The occupancies at the Fermi edge in all non-{\verb!RELAX!} type runs are handled via the tetrahedron method with 
Bl{\"o}chl corrections \cite{Bloechl1994a}. This involves the $N_{KPPRA}$ parameter, as described above. In 
{\verb!RELAX!} type calculations, where the determination of accurate forces is important, some type of 
smearing must be performed. In cases where the material is assumed to be a metal, the Methfessel-Paxton approach 
\cite{Methfessel_prb_1989} is adopted, with a smearing width of 0.10 eV. Gaussian smearing is used in all other types 
of materials, with a smearing width of 0.05 eV.

\subsection{Potentials and basis set} \label{subsec:pseudopot}

The interactions involving the valence electron shells are handled with the potentials provided with the DFT software 
package. In {\small VASP}, these include ultra-soft pseudopotentials (USPP)\cite{Vanderbilt, vasp_JPCM_1994} and 
projector-augmented wavefunction (PAW) potentials\cite{PAW,kresse_vasp_paw}, which are constructed according to the Local 
Density Approximation (LDA)\cite{Ceperley_prl_1980, Perdew_prb_1981}, and the Generalized Gradient Approximation 
(GGA) PW91\cite{VASP_PW91_1,VASP_PW91_2} and PBE\cite{PBE, PBE2} exchange-correlation (XC) functionals. 
The ICSD, Binary Alloy and Heusler databases built according to the {\small AFLOW} standard use the PBE functional combined with 
the PAW potential as the default. The PBE functional is among the best studied GGA functionals used in crystalline systems, while the PAW potentials 
are preferred due to their advantages over the USPP methodology. Nevertheless, defaults have been defined for a number of potential / XC functional 
combinations, and in the case of the Elements database, results are available for LDA, GGA-PW91 and GGA-PBE functionals with both USPP and PAW potentials. 
Additionally, there are a small number of entries in the ICSD and Binary Alloy databases (less than 1\% of the total) which have been calculated with the GGA-PW91 
functional using either the USPP or PAW potential. The exact combination of exchange-correlation functional and potential used for a specific entry
in the {\small AFLOWLIB} database can always be determined by querying the keyword \verb|dft_type| using the {\small AFLOWLIB} REST-API \cite{aflowAPI}.

DFT packages often provide more than one potential of each type per element. The {\small AFLOW}
standardized lists of PAW and USPP potentials are presented in 
Tables \ref{tab:pot_paw} and \ref{tab:pot_uspp}, respectively.
The ``Label'' column in these tables corresponds to the naming convention adopted
by {\small VASP}. The checksum of each file listed in the tables is included in the accompanying supplement 
for verification purposes. 

Each potential provided with the {\small VASP} package has two recommended plane-wave kinetic energy cut-off ($E_{cut}$) 
values, the smaller of which ensures the reliability of a calculation to within a well-defined error. Additionally, 
materials with more than one element type will have two or more sets of recommended $E_{cut}$ values. 
In the {\small AFLOW} standard, the applied $E_{cut}$ value is the largest found among the recommendations for all 
species involved in the calculation, increased by a factor of 1.4.

It is possible to evaluate the the non-local parts of the potentials in real space, rather than in the more computationally 
intensive reciprocal space. This approach is prone to aliasing errors, and requires the optimization of real-space 
projectors if these are to be avoided. The real-space projection scheme is most appropriate for larger systems, e.g. surfaces,
and is therefore not used in the construction of the databases found in the {\small AFLOWLIB} repository.

{\small
  \begingroup
    \begin{table*}
    \centering
      \begin{tabular}[b]{L{1.5cm} L{2.5cm} L{1.5cm} L{2.5cm} L{1.5cm} L{1.0cm}}
        \hline
        \hline
        Element & Label  & Element       & Label      & Element       & Label  \\
        \hline
        H       & H      & Se            & Se         & Gd $\ddagger$ & Gd\_3  \\ 
        He      & He     & Br            & Br         & Tb            & Tb\_3  \\ 
        Li      & Li\_sv & Kr            & Kr         & Dy            & Dy\_3  \\ 
        Be      & Be\_sv & Rb            & Rb\_sv     & Ho            & Ho\_3  \\ 
        B       & B\_h   & Sr            & Sr\_sv     & Er            & Er\_3  \\ 
        C       & C      & Y             & Y\_sv      & Tm            & Tm     \\ 
        N       & N      & Zr            & Zr\_sv     & Yb            & Yb     \\ 
        O       & O      & Nb            & Nb\_sv     & Lu            & Lu     \\ 
        F       & F      & Mo            & Mo\_pv     & Hf            & Hf     \\ 
        Ne      & Ne     & Tc            & Tc\_pv     & Ta            & Ta\_pv \\ 
        Na      & Na\_pv & Ru            & Ru\_pv     & W             & W\_pv  \\ 
        Mg      & Mg\_pv & Rh            & Rh\_pv     & Re            & Re\_pv \\ 
        Al      & Al     & Pd            & Pd\_pv     & Os            & Os\_pv \\ 
        Si      & Si     & Ag            & Ag         & Ir            & Ir     \\ 
        P       & P      & Cd            & Cd         & Pt            & Pt     \\ 
        S       & S      & In            & In\_d      & Au            & Au     \\ 
        Cl      & Cl     & Sn            & Sn         & Hg            & Hg     \\ 
        Ar      & Ar     & Sb            & Sb         & Tl            & Tl\_d  \\ 
        K       & K\_sv  & Te            & Te         & Pb            & Pb\_d  \\ 
        Ca      & Ca\_sv & I             & I          & Bi            & Bi\_d  \\ 
        Sc      & Sc\_sv & Xe            & Xe         & Po            & Po     \\ 
        Ti      & Ti\_sv & Cs            & Cs\_sv     & At            & At     \\ 
        V       & V\_sv  & Ba            & Ba\_sv     & Rn            & Rn     \\ 
        Cr      & Cr\_pv & La            & La         & Fr            & Fr     \\ 
        Mn      & Mn\_pv & Ce            & Ce         & Ra            & Ra     \\ 
        Fe      & Fe\_pv & Pr            & Pr         & Ac            & Ac     \\ 
        Co      & Co     & Nd            & Nd         & Th            & Th\_s  \\ 
        Ni      & Ni\_pv & Pm            & Pm         & Pa            & Pa     \\ 
        Cu      & Cu\_pv & Sm $\dagger$  & Sm         & U             & U      \\ 
        Zn      & Zn     & Sm $\ddagger$ & Sm\_3      & Np            & Np\_s  \\ 
        Ga      & Ga\_h  & Eu            & Eu         & Pu            & Pu\_s  \\ 
        As      & As     & Gd $\dagger$  & Gd         &               &        \\ 
        \hline
        \hline
      \end{tabular}
      \caption{Projector-Augmented Wavefunction (PAW) potentials, parameterized for the LDA, PW91, and PBE 
      functionals, included in the {\small AFLOW} standard.The PAW-PBE combination is used 
        as the default for ICSD, Binary Alloy and Heusler databases. \\
      {\footnotesize $\dagger$: PBE potentials only.}
      {\footnotesize $\ddagger$: LDA and PW91 potentials only.}}
      \label{tab:pot_paw}
    \end{table*}
  \endgroup
}

{\small
  \begingroup
    \begin{table*}
    \centering
      \begin{tabular}[b]{L{1.5cm} L{2.5cm} L{1.5cm} L{2.5cm} L{1.5cm} L{1.0cm}}
        \hline
        \hline
        Element & Label   & Element & Label  & Element & Label \\
        \hline
        H       & H\_soft & As      & As     & Tb      & Tb\_3 \\
        He      & He      & Se      & Se     & Dy      & Dy\_3 \\
        Li      & Li\_pv  & Br      & Br     & Ho      & Ho\_3 \\
        Be      & Be      & Kr      & Kr     & Er      & Er\_3 \\
        B       & B       & Rb      & Rb\_pv & Tm      & Tm    \\
        C       & C       & Sr      & Sr\_pv & Yb      & Yb    \\
        N       & N       & Y       & Y\_pv  & Lu      & Lu    \\
        O       & O       & Zr      & Zr\_pv & Hf      & Hf    \\
        F       & F       & Nb      & Nb\_pv & Ta      & Ta    \\
        Ne      & Ne      & Mo      & Mo\_pv & W       & W     \\
        Na      & Na\_pv  & Tc      & Tc     & Re      & Re    \\
        Mg      & Mg\_pv  & Ru      & Ru     & Os      & Os    \\
        Al      & Al      & Rh      & Rh     & Ir      & Ir    \\
        Si      & Si      & Pd      & Pd     & Pt      & Pt    \\
        P       & P       & Ag      & Ag     & Au      & Au    \\
        S       & S       & Cd      & Cd     & Hg      & Hg    \\
        Cl      & Cl      & In      & In\_d  & Tl      & Tl\_d \\
        Ar      & Ar      & Sn      & Sn     & Pb      & Pb    \\
        K       & K\_pv   & Sb      & Sb     & Bi      & Bi    \\
        Ca      & Ca\_pv  & Te      & Te     & Po      & Po    \\
        Sc      & Sc\_pv  & I       & I      & At      & At    \\
        Ti      & Ti\_pv  & Xe      & Xe     & Rn      & Rn    \\
        V       & V\_pv   & Cs      & Cs\_pv & Fr      & Fr    \\
        Cr      & Cr      & Ba      & Ba\_pv & Ra      & Ra    \\
        Mn      & Mn      & La      & La     & Ac      & Ac    \\
        Fe      & Fe      & Ce      & Ce     & Th      & Th\_s \\
        Co      & Co      & Pr      & Pr     & Pa      & Pa    \\
        Ni      & Ni      & Nd      & Nd     & U       & U     \\
        Cu      & Cu      & Pm      & Pm     & Np      & Np\_s \\
        Zn      & Zn      & Sm      & Sm\_3  & Pu      & Pu\_s \\
        Ga      & Ga\_d   & Eu      & Eu     &         &       \\
        Ge      & Ge      & Gd      & Gd     &         &       \\
        \hline
        \hline
      \end{tabular}
      \caption{Ultra-Soft Pseudopotentials (USPP), parameterized for
        the LDA and PW91 functionals, included in the {\small AFLOW} standard.}
      \label{tab:pot_uspp}
    \end{table*}
  \endgroup
}

\subsection{Fourier transform meshes} \label{subsec:fftmesh}

As mentioned previously, it is not practical to describe the precise default settings that are applied by the {\small AFLOW} 
standard in the specification of the Fourier transform meshes. We
shall just note that they are defined in terms of the grid 
spacing along each of the reciprocal lattice vectors, $\vec{b}_i$. These are obtained from the set of real space lattice 
vectors, $\vec{a}_i$, via $ [\vec{b}_1 \vec{b}_2 \vec{b}_3]^T = 2 \pi [\vec{a}_1 \vec{a}_2 \vec{a}_3]^{-1} $. A distance 
in reciprocal space is then defined by $d_i={\|\vec{b{_i}}\|} /n_i$, where the set of $n_i$ are the number 
of grid points along each reciprocal lattice vector, and where the total number of points in the simulation is 
$n_1 \times n_2 \times n_3$.

The {\small VASP} package relies primarily on the so-called {\textit{dual grid technique}}, which consists of two overlapping 
meshes with different coarseness. The least dense of the two is directly dependent on the applied plane-wave basis, $E_{cut}$, 
while the second is a finer mesh onto which the charge density is mapped. The {\small AFLOW} standard relies on placing 
sufficient points in the finer mesh such that wrap-around ("aliasing") errors are avoided. In terms of the quantity $d_i$, 
defined above, the finer grid is characterized by $d_i \approx 0.10${\textit{ \r{A}}$^{-1}$}, while the coarse grid results 
in $d_i \approx 0.15${\textit{ \r{A}}$^{-1}$}. These two values are approximate, as there is significant dispersion in 
these quantities across the various databases.

\subsection{DFT+U corrections} \label{subsec:Hubbard}

Extended systems containing {\textit d} and {\textit f} block elements are often poorly represented within DFT due to 
the well known self interaction error (SIE)\cite{Perdew_prb_1981}. The influence that the SIE has on the energy gap of 
insulators has long been recognized, and several methods that account for it are available. These include the 
{\textit{GW}} approximation\cite{Hedin_GW_1965}, the rotationally invariant approach introduced by 
Dudarev\cite{Dudarev_dftu} and Liechtenstein\cite{Liechtenstein1995} (denoted here as DFT+U), as well as the recently 
developed ACBN0 pseudo-hybrid density functional\cite{curtarolo:art93}.

The DFT+U approach is currently the best suited for high-throughput investigations, and is therefore included in 
the {\small AFLOW} standard for the entire ICSD database, and is also used for certain entries in the Heusler 
database containing the elements O, S, Se, and F. It is not used for the Binary Alloy database.
This method has a significant dependence on parameters, as each atom is associated with 
two numbers, the screened Coulomb parameter, $U$, and the Stoner exchange parameter, $J$. These are usually reported 
as a single factor, combined via $U_{\text{eff}}=U-J$. The set of $U_{\text{eff}}$ values associated with the 
{\textit d} block elements\cite{curtarolo:art58,curtarolo:art68} are presented in Table \ref{tab:Ud}, to which the 
elements In and Sn have been added.

A subset of the {\textit f}-block elements can be found among the systems included in 
the {\small AFLOWLIB} consortium databases. We are not aware of the existence of a systematic search for the best set 
of $U$ and $J$ parameters for this region of the periodic table, so we have relied on an in-house 
parameterization\cite{curtarolo:art58} in the construction of the databases. The values used are reproduced 
in Table \ref{tab:Uf}. Note that by construction the SIE correction must be applied to a pre-selected value of the 
$\ell$-quantum number, and all elements listed in Table \ref{tab:Ud} correspond to $\ell=2$, while those
found in Table \ref{tab:Uf} correspond to $\ell=3$.

{\small
   \begin{table}[h]
     \centering
     \begin{tabular}[b]{L{1.5cm} C{2.5cm} L{1.5cm} C{1.0cm}}
       \hline
       \hline
       Element & $U_{\text{eff}}$ & Element & $U_{\text{eff}}$ \\
       \hline
       Sc \cite{ScUJ} & 2.9 & W  \cite{NbUJ} & 2.2 \\
       Ti \cite{TiUJ} & 4.4 & Tc \cite{NbUJ} & 2.7 \\
       V  \cite{VUJ}  & 2.7 & Ru \cite{NbUJ} & 3.0 \\
       Cr \cite{CrUJ} & 3.5 & Rh \cite{NbUJ} & 3.3 \\
       Mn \cite{CrUJ} & 4.0 & Pd \cite{NbUJ} & 3.6 \\
       Fe \cite{FeUJ} & 4.6 & Ag \cite{AgUJ} & 5.8 \\
       Co \cite{VUJ}  & 5.0 & Cd \cite{ZnUJ} & 2.1 \\
       Ni \cite{VUJ}  & 5.1 & In \cite{ZnUJ} & 1.9 \\
       Cu \cite{CrUJ} & 4.0 & Sn \cite{SnUJ} & 3.5 \\
       Zn \cite{ZnUJ} & 7.5 & Ta \cite{NbUJ} & 2.0 \\
       Ga \cite{GaUJ} & 3.9 & Re \cite{NbUJ} & 2.4 \\
       Sn \cite{SnUJ} & 3.5 & Os \cite{NbUJ} & 2.6 \\
       Nb \cite{NbUJ} & 2.1 & Ir \cite{NbUJ} & 2.8 \\
       Mo \cite{NbUJ} & 2.4 & Pt \cite{NbUJ} & 3.0 \\
       Ta \cite{SnUJ} & 2.0 & Au & 4.0 \\
       \hline
       \hline
     \end{tabular}
     \caption{$U_{\text{eff}}$ parameters applied to {\textit d} orbitals.}
     \label{tab:Ud}
   \end{table}
}

{\small
   \begin{table}[h]
     \centering
     \begin{tabular}[b]{L{1.5cm} C{1.5cm} C{1.5cm} L{1.5cm} C{1.5cm} C{1.5cm}}
       \hline
       \hline
       Element & {\textit U} & {\textit J} & Element & {\textit U} & {\textit J} \\
       \hline
       La \cite{LaUJ} & 8.1 & 0.6 & Dy \cite{DyUJ} & 5.6 & 0.0  \\
       Ce \cite{CeUJ} & 7.0 & 0.7 & Tm \cite{TmUJ} & 7.0 & 1.0  \\
       Pr \cite{PrUJ} & 6.5 & 1.0 & Yb \cite{YbUJ} & 7.0 & 0.67 \\
       Nd \cite{aflowSCINT}& 7.2 & 1.0 & Lu \cite{LaUJ} & 4.8 & 0.95 \\
       Sm \cite{aflowSCINT}& 7.4 & 1.0 & Th \cite{ThUJ} & 5.0 & 0.0  \\
       Eu \cite{aflowSCINT}& 6.4 & 1.0 & U  \cite{UUJ}  & 4.0 & 0.0  \\
       Gd \cite{GdUJ} & 6.7 & 0.1 &    &     &      \\
       \hline
       \hline
     \end{tabular}
     \caption{$U$ and $J$ parameters applied to selected {\textit f}-block elements.}
     \label{tab:Uf}
   \end{table}
}

\subsection{Spin polarization} \label{subsec:SpinPol}

The first of the two {\verb!RELAX!} calculations is always performed in a collinear spin-polarized fashion. 
The initial magnetic moments in this step are set to the number of atoms in the system, e.g. 1.0 $\mu B/$atom. If 
the magnetization resulting from the {\verb!RELAX1!} step is found to be below 0.025 $\mu B/$atom, {\small AFLOW}
economizes computational resources by turning spin polarization off in all ensuing calculations. Spin-orbit coupling
is not used in the current {\small AFLOW} standard, since it is still
too expensive to include in a HT framework.

\subsection{Calculation methods and Convergence criteria} \label{subsec:convergence}

Two nested loops are involved in the DFT calculations used by {\small AFLOW} in the construction of the databases.
The inner loop contains routines that iteratively optimize the electronic degrees of freedom (EDOF), and features 
a number of algorithms that are concerned with diagonalizing the Kohn-Sham (KS) Hamiltonian at each iteration. 
The outer loop performs adjustments to the system geometry (ionic degrees of freedom, IDOF) until the forces acting 
on the system are minimized.

The convergence condition for each loop has been defined in terms of an energy difference, $\delta E$. If successive 
energies resulting from the completion of a loop are denoted as $E_{i-1}$ and $E_i$, then 
convergence is met when the condition $\delta E \geqslant E_i - E_{i-1}$ is fulfilled. Note that $E_i$ can either be 
the electronic energy resulting from the inner loop, or the configurational energy resulting from the outer loop. 
The electronic convergence criteria will be denoted  as $\delta E_{elec}$, and the ionic criteria as $\delta E_{ion}$. 
The {\small AFLOW} standard relies on $\delta E_{elec} = 10^{-5}$ eV and $\delta E_{ion} = 10^{-4}$ eV for entries in the 
Elements database. All other databases include calculations performed with $\delta E_{elec} = 10^{-3}$ eV and $\delta E_{ion} = 10^{-2}$ eV.

Optimizations of the EDOF depend on sets of parameters that fall under three general themes: initial guesses, diagonalization 
methods, and charge mixing. The outer loop (optimizations of the IDOF) is concerned with the lattice vectors and the ionic
positions, and is not as dependent on user input as the inner
loops. These are described in the following paragraphs.

\subsubsection{Electronic degrees of freedom} \label{subsec:econverge}

The first step in the process of optimizing the EDOF consists of choosing a trial charge density and a trial 
wavefunction. In the case of the non-{\verb!BANDS!}-type calculations, the trial wavefunctions are initialized 
using random numbers, while the trial charge density is obtained from the superposition 
of atomic charge densities. The {\verb!BANDS!} calculations are not self-consistent, and thus do not feature 
a charge density optimization. In these cases the charge density obtained from the previously performed {\verb!STATIC!} 
calculation is used in the generation of the starting wavefunctions.

Two iterative methods are used for diagonalizing the KS Hamiltonian: the Davidson blocked scheme 
(DBS)\cite{Liu_rep_1978,Davidson_1983}, and the preconditioned residual minimization method -- direct inversion in 
the iterative subspace (RMM--DIIS)\cite{VASP4_4}. Of the two, DBS is known to be the slower and more stable option. 
Additionally, the subspace rotation matrix is always optimized. These methods are applied in a manner that is dependent 
on the calculation type:

\begin{enumerate}[i.]
  \item {\verb!RELAX!} calculations. Geometry optimizations contain at least one determination of the system 
  forces. The initial determination consists of 5 initial DBS steps,
  followed by as many RMM-DIIS steps as needed to
  fulfill the $\delta E_{elec}$ condition. Later determinations of
  system forces are performed by a similar
  sequence, but only a single DBS step is applied at the outset of the process. Across all 
  databases the minimum of number of electronic iterations for {\verb!RELAX!} calculations is 2. The maximum number is set
  to 120 for entries in the ICSD, and 60 for all others. 
  \item non-{\verb!RELAX!} calculations. In {\verb!STATIC!} or 
  {\verb!BANDS!} calculations, the diagonalizations are always performed using RMM--DIIS. The minimum number of electronic
  iterations performed during non-{\verb!RELAX!} calculations is 2, and the maximum is 120.
\end{enumerate}

If the number of iterations in the inner loop somehow exceed the limits listed above, the calculation breaks 
out of this loop, and the system forces and energy are determined. If the $\delta E_{ion}$ convergence condition is 
not met the calculation re-enters the inner loop, and proceeds normally.

Charge mixing is performed via Pulay's method\cite{Pulay_cpl_1980}. The implementation of this charge mixing 
approach in the {\small VASP} package depends on a series of parameters, of which all but the maximum $\ell$-quantum number 
handled by the mixer have been left in their default state. This parameter is modified 
only in systems included in the ICSD database which contain the elements 
listed in Tables \ref{tab:Ud} and \ref{tab:Uf}. In practical terms, the value applied in these cases is the maximum 
$\ell$-quantum number found in the PAW potential, multiplied by 2.

\subsubsection{Ionic degrees of freedom and lattice vectors} \label{subsec:iconverge}

The {\verb!RELAX!} calculation type contains determinations of the forces acting on the ions, as well as the full system
stress tensor. The applied algorithm is the conjugate gradients (CG) approach\cite{press1992numerical}, which depends on 
these quantities for the full optimization of the system geometry, i.e. the ionic positions, the lattice vectors, as well 
as modifications of the cell volume. The implementation of CG in {\small VASP} requires minimal 
user input, where the only independent parameter is the initial scaling factor which is always left at its
default value. Convergence of the IDOF, as stated above, depends on the value for the $\delta E_{ion}$ parameter, 
as applied across the various databases. The adopted $E_{cut}$ (see discussion on ``Potentials and basis set'', 
section \ref{subsec:pseudopot}) makes corrections for Pulay stresses unnecessary.

Forces acting on the ions and stress tensor are subjected to Harris-Foulkes\cite{Harris_prb_1985} corrections. 
Molecular dynamics based relaxations are not performed in the construction of the databases found in the 
{\small AFLOWLIB} repository, so any related settings are not applicable to this work.

\subsection{Output options} \label{subsec:output}

The reproduction of the results presented on the {\small AFLOWLIB} website also depends on a select few parameters that
govern the output of the DFT package. The density of states plots are generated from the {\verb!STATIC!} 
calculation. States are plotted with a range of -30 eV to 45 eV, and with a resolution of 5000 points. The band 
structures are plotted according to the paths of {\textit k}-points generated for a {\verb!BANDS!} 
calculation\cite{curtarolo:art58}. All bands found between -10 eV and 10 eV are included in the plots.

\section{Conclusion} \label{sec:conclusion}

The {\small AFLOW} standard described here has been applied in the automated creation of the {\small AFLOWLIB} database of 
material properties in a consistent and reproducible manner. The use of standardized parameter sets facilitates 
the direct comparison of properties between different materials, so that specific trends can be identified to assist 
in the formulation of design rules for accelerated materials development. Following this {\small AFLOW} standard should 
allow materials science researchers to reproduce the results reported by the {\small AFLOWLIB} consortium, as well as to
extend on the database and make meaningful comparisons with their own results.  

\section{Acknowledgments} \label{sec:acknowledgments}

We thank Dr. Kesong Yang for various technical discussions. We would like to acknowledge support by the 
by DOD-ONR (N00014-13-1-0635, N00014-11-1-0136, N00014-09-1-0921), and CT, JJP and SC acknowledge support from the DOE 
(DE-AC02- 05CH11231), specifically the Basic Energy Sciences program under Grant \# EDCBEE. The 
{\small AFLOWLIB} consortium would like to acknowledge the Duke University Center for Materials Genomics and the CRAY 
corporation for computational support.

\newpage

\bibliographystyle{elsarticle-num-names}
\bibliography{xcamilo,xcormac,xstefano4}
\end{document}